\documentstyle[epsfig,12pt,epsf,cite]{article}
\begin{document}

\begin{center}
{\large\bf Formation of $\sigma$ Mesic Nuclei\\
 in (d,t) and (d,$^3$He) Reactions
   }\\ 	[3mm]

{S.~Hirenzaki$^a$, H.~Nagahiro$^a$\footnote{e-mail: deko@phys.nara-wu.ac.jp,
address: Kitauoya-Nishi-machi, Nara, 630-8506, Japan.}, T.~Hatsuda$^b$ and  T.~Kunihiro$^c$
}\\[3mm]

{\small \em
$^a$Department of Physics, Nara Women's Univ., \\ Nara 630-8506,
Japan\\
$^b$Department of Physics, Univ. of Tokyo, \\ Tokyo 113-0033,Japan\\
$^c$Yukawa Institute for Theoretical Physics,
Kyoto University,\\ Kyoto 606-8502, Japan}\\ [3mm]
\end{center}

\abstract{
We explore the possibility to observe an enhanced 2$\pi$ correlation
in the scalar iso-scalar ($\sigma$)  channel in nuclei by
(d,t) and (d,$^3$He) reactions
to obtain new information on the partial restoration of
chiral symmetry in the nuclear medium.  The sensitivity of the reaction
spectra on the change of the chiral condensate in the nuclear medium
 is  examined. 
}

\section{Introduction}

 Mesic atoms and mesic nuclei are the useful laboratory for
 studying the meson-baryon interactions and the meson
  properties in the nuclear medium.
 Among others, pionic atoms have been long utilized
for such studies \cite{ericson88}.
The standard method to produce   pionic atoms
starts with
injecting slow negative pions into matter:  The  pions
will be stopped and  trapped in outermost orbits of atoms,
 then lose
 energy by emitting Auge electrons and x-rays to cascade down to
deeper atomic states.
The binding energies and widths
of the pionic bound states provide us with a unique
 information of the pion-nucleus interaction \cite{laat91}.

 However,  the x-ray cascade ceases at the so called
 'last orbital', where the pions are absorbed by the nucleus without
going into deeper atomic states. This prevents us from studying
 the deep pionic states  that
 have large overlap with nuclear density \cite{toki89}.

 To overcome this difficulty,
 the recoilless meson production by the
(d,$^3$He) reaction has been proposed \cite{hirenzaki91};
this method tries to directly create deeply bound pions
by the nuclear reaction without the cascade down of the pion
 from the outer most orbits.
This direct method was proved to be
effective and  powerful for the study of
far deeply bound pionic atoms \cite{yamazaki96,gillitzer00}:
We can now determine the binding energies and widths of
the deepest bound states such as 1s and 2p atomic states
 from the energy spectra of the emitted $^3$He.
 We remark that the (d,$^3$He)-reaction is so powerful that
it may be applied to the creation of
even neutral meson-nucleus bound states  such as  the $\eta$ and
$\omega$ mesic nuclei \cite{hayano99}.

 In this paper,
  we explore  a possibility of  producing  $\sigma$ mesic nuclei
by the one-nucleon pick-up reactions.
 The  $\sigma$ meson (or $f_0(400-1200)$),
  which is a scalar-isoscalar (I=J=0) resonance,
 is considered to be the chiral partner of the pion and
 is related to the amplitude fluctuation of the chiral condensate
 $\langle \bar{q}q \rangle$ in the QCD vacuum \cite{HK94}.
 In the free space, $\sigma$
may appear as a broad resonance at best,  since it has a
 large  decay width into two pions \cite{pdg98,sigma-w}.
 In other words, the physical $\sigma$ is  a superposition of
 bare-$\sigma$ and 2$\pi$.  On the other hand, the $\sigma$
  may appear as a sharp resonance at finite temperature and/or
  baryon density, if  the partial restoration of chiral symmetry
takes place.  This is because the partial degeneracy of
 $\sigma$ and $\pi$ in the medium leads to the
  suppression of the phase space of the decay
  $\sigma\rightarrow$2$\pi$  \cite{hatsuda84}.
  In this context,
  the spectral function in the
 scalar-isoscalar channel  at finite temperature
 ($T$)  has been studied in the linear $\sigma$-model, and
  the significant spectral
  softening  and the spectral enhancement
   near the 2$\pi$ threshold have been proposed as a characteristic
 signal of the partial restoration of chiral symmetry
 \cite{CH98}.
 The idea was also generalized to the scalar-isoscalar
 spectral-function in nuclear
  matter on the basis of the linear and non-linear $\sigma$-model,
 and a similar softening and enhancement have shown to take
 place even around nuclear matter density  \cite{hatsuda99}.
  This phenomenon may be also related to
 the data reported by  CHAOS collaboration
 and by Cristal Ball collaboration, in which the
 softening of the invariant mass
 distribution of emitted 2$\pi$
 in A($\pi$,$\pi\pi$)A$^\prime$ reaction is suggested
 \cite{CHAOS,CB,TOT}. Intensive theoretical investigations
 are now under way to see whether those data
 supply us with an information of the partial restoration of
 chiral symmetry in nuclear matter \cite{shuck,paris01}.

 A unique feature of the $\sigma$ meson is that the mass
and width are smaller  in the medium than in the vacuum.
It implies that there is a chance to create
 the $\sigma$ mesic nuclei with a relatively small width,
 if one can produce  deeply bound states in heavy nuclei.
 Finding such states will provide us with more information on
the chiral properties of the nuclear medium and how
 partial restoration of  chiral symmetry occurs in the
medium. In fact, some experiments to produce the sigma meson in a nucleus
 to explore the possible restoration in nuclear medium
 were proposed some time
 ago\cite{kuni95}.

To maximize the density effect in the $\sigma$ formation reactions,
heavier nuclear targets should be preferable.
 As the actual reaction to create the $\sigma$ mesic nuclei,
 the one-{\em neutron} pick-up (d,t) reaction is
 appropriate from the theoretical point of view, because
 the heavy nuclei are neutron-rich.
 On the other hand, from the experimental point of view,
 the deuteron and the triton
 have the same magnetic rigidity
 in the recoilless kinematics, which
 causes difficulty in experimental isolation of the
 triton from the deuteron in the final state.
 The one-proton pick-up (d,$^3$He) reaction, which
 was proved useful to produce the deeply bound pionic atoms,
 does not have this difficulty since the
 magnetic rigidity of $d$ and $^3$He are different.
 Because of these reasons,
  we consider both the (d,t) and (d,$^3$He) reactions
 in this paper and
 study the formation cross sections of the $\sigma$ mesic nucleus
 in these reactions.
\footnote{
  For the formation of the $\eta$ and $\omega$ mesic nuclei \cite{hayano99},
 light nuclei are chosen as targets to avoid complex subcomponents of
 expected spectra, and  the one-proton pick-up (d,$^3$He)
 reaction is chosen to
 avoid the background from the incident deuteron.}

In section 2, we calculate the spectrum in the $\sigma$ channel in
 nuclei.  We show the expected spectra in (d,t) and (d,$^3$He) reactions in
section 3.
Section 4 is devoted to a summary and concluding remarks.

%
%
%
%
%
\section{Spectrum in  the $\sigma$ channel in nuclei}

In order to obtain the $\sigma$ spectrum in finite nuclei, we
 start with the Klein-Gordon equation with the $\sigma$ self-energy
 $\Sigma_{\sigma}$ which
 contains both the $\sigma \rightarrow 2 \pi$ process and the
 interactions with the nucleons,
\begin{equation}
[ - \vec{\nabla}^2 + m_{\sigma}^2 + \Sigma_{\sigma}(\omega,\rho(\vec{r}))]
\phi_{\sigma}(\vec{r}) =
\omega^2 \phi_{\sigma}(\vec{r}) ,
\label{eqn:KG}
\end{equation}
where $\rho(\vec{r})$ is the local nuclear density,
 $m_{\sigma}$ is the bare mass of the $\sigma$-meson,
 and $\omega$ is the energy eigenvalue.
 Here we have neglected the momentum dependence of the self-energy.
 From the solution of this equation,
 we obtain the complex energy eigenvalues of the $\sigma$ states in
nuclei.  Since the self-energy is a
 function of $\omega$ carried
by the $\sigma$ meson, we need to
 solve  the above Klein-Gordon equation  self-consistently.
We notice that the eigenstates obtained here are not orthogonal to each
other due to the energy dependence of the self-energy. In this paper, we
neglect this point and assume the orthogonality of the states.

The  self-energy $\Sigma_{\sigma} (\omega, \rho)$
 for uniform nuclear matter is evaluated in Ref.
\cite{hatsuda99} based on the SU(2) linear $\sigma$ model in the
one-loop approximation. The explicit form reads
\begin{eqnarray}
{\rm Re} \Sigma_{\sigma}(\omega; \rho) & = &
      - \lambda \sigma_0^2 \ \left( 1 - \Phi(\rho) \right)  \nonumber \\
 &  -  & {\lambda \over 32 \pi^2} \  [
          \ \
               m_{\pi}^2 \ (1- \ln {m_{\pi}^2 \over \kappa^2})
              +m_{\sigma}^2 \ (1- \ln {m_{\sigma}^2 \over \kappa^2})
         \nonumber \\
 &  +  &  {1 \over 3} \lambda \sigma_0^2 \  (Q_{\pi} +
              2- \ln {m_{\pi}^2 \over \kappa^2}) \nonumber \\
  &  +  & \lambda \sigma_0^2 \ (Q_{\sigma} +
              2- \ln {m_{\sigma}^2 \over \kappa^2}) \ \ \  ] , \\
{\rm Im} \Sigma_{\sigma}(\omega; \rho) & = &
  - {\lambda^2 \over 32 \pi}  \sigma_0^2 \
  \left[
    {1  \over 3}   P_{\pi} \theta (\omega - 2 m_{\pi} )
    +  P_{\sigma} \theta (\omega - 2 m_{\sigma} ) \right] ,
\end{eqnarray}
where
\begin{eqnarray}
Q_{\varphi} & = & P_{\varphi} \ \ln \ {1-P_{\varphi} \over
                     1+ P_{\varphi} } \ \ \ ({\rm for} \ \
                     2m_{\varphi} \le \omega ) \\
               & = & -2 P_{\varphi} \ {\rm arctan} \ {1 \over
                    P_{\varphi}} \ \ \ ({\rm for} \ \
                     \omega \le 2m_{\varphi}  ) ,
\end{eqnarray}
with $P_{\varphi}  =  \left| 1 - 4 m_{\varphi}^2 / \omega^2 \right| ^{1/2}$.
 Here,  $\varphi$ denotes either $\pi$ or $\sigma$.
 $\kappa$ is a renormalization point
 in the minimal subtraction scheme.
  $\Phi(\rho)$ is defined to express the chiral condensate in nuclear
matter $\langle\sigma\rangle_{\rho}$ as
$\langle\sigma\rangle_{\rho} = \sigma_0 \Phi (\rho)$.
In the linear density approximation,
\begin{eqnarray}
\Phi(\rho)=1-C{ \rho \over \rho_0},
\end{eqnarray}
with $C$ being the measure of the
 rate of how strongly chiral symmetry is restored in the nuclear medium.
 Depending on how $\langle\sigma\rangle$ is related to
 $\langle \bar{q}q \rangle$, $C$ may take a value 0.1 -0.4;
see the second reference of
\cite{HK94}.

The first term in Re$\Sigma_{\sigma} $
  corresponds to the mean field correction,
$\Sigma_{MF}(\rho)=\lambda \sigma_0 (\langle\sigma\rangle_{\rho} - \sigma_0)
= - \lambda \sigma_0^2 (1 - \Phi (\rho)) $ which
is responsible for the $\rho$ dependence of the self-energy and provide
the strong attractive potential for the $\sigma$ meson
 in heavy nuclei.  Due to this
attraction,  bound $\sigma$ states may be formed.
In order to construct
the $\sigma$-nucleus optical potential, we use
the local density approximation and consider the density $\rho$ to be the
nuclear density distribution $\rho (\vec{r})$.
The parameters used in this paper are obtained
from the first reference in  \cite{CH98} and
listed in Table 1.
In principle, we can choose any value of the renormalization point
$\kappa$ since the value of coupling constant $\lambda$ depends on
$\kappa$ and the simultaneous change of $\kappa$ and $\lambda$ 
makes no differences to the physical results. However, due to the 1-loop
approximation in our formula, we have a weak $\kappa$ dependence of the
physical results and fix $\kappa$ to a certain value so as to reproduce
the vacuum observables \cite{CH98}.

 We show in Fig.1
the attractive
potential in the nuclear medium $V(r) \equiv \Sigma_{MF}(\rho(r)) / 2m_{\sigma}$
for $^{208}$Pb, where the  density
distribution is assumed to be
of Woods-Saxon type with 6.5 [fm] and 0.5 [fm] for the radius and diffuseness
parameters, respectively.   The three curves correspond to
C=0.2, 0.3 and 0.4, respectively.
 Im$\Sigma_{\sigma}$ is independent of the nuclear
density in the present (one-loop) approximation.

We show in Fig. 2 the bound state spectra of the $\sigma$ meson in
$^{208}$Pb for three cases with C=0.2, 0.3 and
0.4, together with the widths of the low lying
  states. The width is defined by $\Gamma = - {\rm Im} \Sigma_{\sigma}
 (\omega_{nl}) /m_{\sigma}$ with the real part
of the eigenenergies
$\omega_{nl}$ for $(nl)$ bound states.
We find 17, 26 and 32
bound eigenstates $(nl)$ of the $\sigma$ for C=0.2, 0.3
and 0.4, respectively.
The widths tend to decrease
for deeper bound states because of the suppression of the
 $\pi\pi$ decay phase space.
 Furthermore, for C=0.3 and 0.4,
there appear so deep bound states that they have no width because
of the hindrance of the  decay into  $2 \pi$ in the one-loop level.
We should mention here that the meaning of the eigenstates with the width
for larger than the separation of the levels is not clear and that the
applicability of the ordinary quantum mechanics to these states should be
discussed somewhere.

In Fig.3 (a) the spectral function of the bound $\sigma$ states
 defined below is shown for $l=0$ states for C=0.3:

\begin{equation}
\rho_{nl} (\omega)= - \frac{1}{\pi} \frac{{\rm Im} \Sigma_{\sigma}
}{(\omega^2 - \omega_{nl}^2)^2 + ({\rm Im} \Sigma_{\sigma}
)^2},
\label{eqn:rhonl}
\end{equation}

\noindent
 For comparison, we also show,
  in Fig.3(b), the spectral function of $\sigma$ in
 nuclear matter reported in Ref.\cite{hatsuda99}, which is defined as;
\begin{equation}
\rho (\omega)= - \frac{1}{\pi} \frac{{\rm Im} \Sigma_{\sigma}
}{(\omega^2 - m_{\sigma}^2 - {\rm Re} \Sigma_{\sigma})^2 + ({\rm Im}
\Sigma_{\sigma}
)^2}.
\label{eqn:rhomat}
\end{equation}
In Eq.(\ref{eqn:rhonl}), we replace  $m_{\sigma}^2 + {\rm Re}
\Sigma_{\sigma}$ in Eq.(\ref{eqn:rhomat}) into $\omega_{nl}^2$ for the bound
eigenstate ($nl$) of the $\sigma$ in the nucleus.
 There exists a significant
enhancement of the spectral function at $\omega \sim 2 m_{\pi}$ for
deeply bound states.
In Eq.(\ref{eqn:rhonl}) and Fig.3(a), we added 5MeV to the
Im$\Sigma_\sigma$
in order to regularize the infinity at $\omega=\omega_{nl}$ for
Im$\Sigma_\sigma=0$
eigenstates and to depict the spectral function in the figure.

The full  spectral function of the $\sigma$
 in $^{208}$Pb may be given by
\begin{equation}
\label{simple-sum}
\rho_{\sigma-Pb}^{full}(\omega) = \sum_{nl} (2l+1) \rho_{nl} (\omega),
\end{equation}
i.e., a simple superposition of all
 possible bound states with degeneracy factor $(2l + 1)$;
here the possible non-orthogonality of the eigenstates is assumed small,
which should be checked in a future work.
The numerical results of $\rho_{\sigma-Pb}^{full}(\omega)$
are shown  in Fig.4 for C=0.2, 0.3 and 0.4.
  The enhanced peak of the
 spectral function due to the deep bound states of
  $^{208}$Pb seen in Fig.3(a) is now somewhat spread
  due to large contributions from the shallow states
  whose spectral functions are close to those  in the vacuum.
 Nevertheless, there still remains
   a significant enhancement  near the 2$\pi$ threshold
  for C=0.3 and 0.4 since they have bound states below
 2$m_{\pi}$.

\section{(d,t) and (d,$^3$He) spectra in the  $\sigma$ channel}

In the realistic experiments using nuclear reactions,
the contribution of each bound state to the cross section is
 not equal  but their relative significance is strongly
dependent on the reaction mechanism.
 In order to take account of
the reaction mechanism into the expected shape of the spectral function,
we introduce the effective number $N_{\rm eff}$
as the relative
weight of each bound state \cite{hirenzaki91}.

Before proceedings to the calculation, we give
a detailed argument on the  effectiveness of the (d, t) reactions
 for our purpose.
 First of all, deeper bound $\sigma$ states are
 preferable to observe a
significant change of the spectral function: In fact,
 we have seen in section 2 that
the $\sigma$ mesic $s$ and $p$ states have large binding energies and  small
widths.
Furthermore,  the substitutional states are largely enhanced in the recoilless
 reactions, and it is important to select proper configurations of meson
 bound states and single nucleon (proton or neutron) states to have large
 signals.
Heavy nuclei, suited to produce deep $\sigma$ states, have different
 single particle configuration for protons and neutrons.
The $\sigma$ mesic $s$ and $p$ states
 will be largely populated in the configurations of [$s^{\sigma}
 \otimes s^{proton}_{1/2}$], [$p^{\sigma} \otimes p^{neutron}_{1/2}$] and
 [$p^{\sigma}  \otimes p^{neutron}_{3/2}$] with proton and neutron levels
 in the valence shell in the Pb region.  Since the number of neutron is larger
 than that of proton in these states, we can expect  larger
 enhancement of the [$p^{\sigma} \otimes p^{neutron}_{1/2}$] and
 [$p^{\sigma}  \otimes p^{neutron}_{3/2}$] configurations and more
 significant changes in the spectral function
 than for [$s^{\sigma}
 \otimes s^{proton}_{1/2}$] in the spectrum.
 Thus,
 the one-neutron pick-up (d,t) reaction
 could be theoretically more effective for the formation of $\sigma$-mesic
 nuclei than the  proton pick-up (d,$^3$He) reaction.
 On the other hand, from the experimental point of view,
 the (d,$^3$He) reaction has less difficulty to
 detect the final state as mentioned in the Introduction.
 Hence, we
 investigate  both (d,t) and (d,$^3$He) reactions.

First, we consider the (d,t) reactions. All formula shown below can also be
applied to the (d,$^3$He) cases similarly.
Total spectrum for the (d,t) reaction is  given by

\begin{equation}
\rho_{\sigma-Pb}^{(d,t)}(\omega) = \sum_{nl} N_{\rm eff} \
 \rho_{nl} (\omega).
\label{eqn:totalspect}
\end{equation}

\noindent
Here  $N_{\rm eff}$ is defined by
\begin{eqnarray}
N_{\rm eff} = \sum_{J M m_s}
|\int d^3r
\chi^{\ast}_f({\bf r}) \xi^{\ast}_{1/2,m_s}
[\phi^{\ast}_{l_{\sigma}}({\bf r}) \otimes \psi_{j_n}({\bf r})]_{JM}
\chi_i({\bf r})|^2,
\end{eqnarray}
\noindent
where  $\psi_{j_n}$ and $\phi_{l_{\sigma}}$
 are the neutron and the $\sigma$ wave functions, respectively, while
 $ \xi_{1/2,m_s}$ the spin wave function.

We adopt the
harmonic-oscillator wave function for $\psi_{j_n}$.
We take the spin average with respect to $m_s$ so as to
take into account the possible spin direction of
the neutrons in the target nucleus. $\chi_i$ and $\chi_f$ are the initial
and the
final distorted waves of the
projectile and the ejectile, respectively.

 We use the Eikonal
approximation with $z$ being the coordinate in the
beam direction, then $\chi_f^* \chi_i$ is written as
\begin{eqnarray}
\chi^{\ast}_f({\bf r}) \chi_i({\bf r}) = \exp (i{\bf q \cdot r})D(z, {\bf b}),
\end{eqnarray}
where the distortion factor D(z,{\bf b}) is defined by
\begin{eqnarray}
D(z, {\bf b}) = \exp \left[
-\frac{1}{2} \sigma_{dN} \int^{z}_{-\infty}d z^{\prime}
\rho_A (z^{\prime},{\bf b})
-\frac{1}{2} \sigma_{tN} \int^{\infty}_{z}d z^{\prime}
 \rho_{A-1} (z^{\prime},{\bf b})
\right].
\end{eqnarray}
Here, $\sigma_{dN}$($\sigma_{tN}$)
is  the deuteron-nucleon (triton-nucleon) total cross sections;
$\rho_A(z,{\bf b})$ and $\rho_{A-1}(z,{\bf b})$ are
the density distributions of the target and the daughter nuclei
 with an impact parameter ${\bf b}$, respectively.

The $\sigma$ wave functions $\phi_{l_{\sigma}}$ used to calculate
the effective number $N_{\rm eff}$ are obtained by solving
the Klein-Gordon equation (\ref{eqn:KG}).
 $N_{\rm eff}$ is calculated in exactly the same way as
the pion production
\cite{umemoto00}
at the resonance energy $\omega_{nl}$ and is known to give a  good
account of the relative contributions of bound states for
the pionic atoms\cite{yamazaki96}.

We shall now calculate  N$_{\rm eff}$ of (d,t) and (d,$^3$He) reactions
at T$_d$ = 1.5 and 4
GeV  with Pb being the target.
 At T$_d$ = 1.5 GeV, both reactions satisfy the recoilless
condition and
the momentum transfer $q \equiv |\vec{q} |$
  is nearly equal to zero at $\omega \sim 2 m_\pi $
where the spectral enhancement is expected.
 On the other hand,  at T$_d$ = 4 GeV,
the recoilless condition ($ q \sim$ 0) is satisfied
 at $\omega$ = 500-600 MeV where the spectral
function in the vacuum has the largest value.
In the calculation, we have included
all the single-particle neutron (for the (d,t) reactions) or proton
(for the (d,$^3$He) reactions) states in the valence shell of $^{208}$Pb.

In Fig.5, the calculated total spectral function defined in Eq.
(\ref{eqn:totalspect}) for the (d,t) and (d,$^3$He) reactions are shown.
The total spectral function for the (d,$^3$He) reaction is defined
similarly using the N$_{eff}$ for the (d,$^3$He) reactions and shown
together in Fig. 5.
We find that the peak structure around $\omega\sim2m_\pi$ is suppressed
from the simple sum in Eq.(\ref{simple-sum})
because of the
distortion effects of the deuteron and triton or $^3$He.
Since the contributions from shallow bound states dominate the spectral 
function, all results in Fig. 5 show the similar energy dependence even
with different $C$ values.

 As a general tendency, the magnitude of the $\rho^{(d,t)}_{\sigma -
 Pb}$ is larger than that of $\rho^{(d,^3He)}_{\sigma -
 Pb}$ due to the existence of the larger number of neutrons than protons
 in the target nucleus, and the magnitude of the spectral function with
 larger $C$ value is larger
  because there exist
more $\sigma$ bound states for larger $C$  due to the stronger attractive
optical
potential for the $\sigma$ meson.
However, due to the reaction mechanism, we can find several exceptions of
the general tendencies described above.  For example, the spectrum with
$C$=0.4 for the (d,$^3$He) reaction at T$_d$=4 GeV is smaller than other
cases
with $C=0.2$ and $0.3$. This is due to the lack of shallow $\sigma$
bound states with ${\rm B.E.}\leq 20\sim30$ MeV and $l=0, 2, 4$, which
are expected to provide large contributions to $\rho$.

 Fig.5 shows a sharp peak near 2$\pi$ threshold for $C=0.4$ with
 T$_d$=1.5 GeV. This
 originates from
 the deep bound states in the case of $C$=0.4.
In contrast, the sharp peak disappears
 at T$_d$=4 GeV; this is because the recoilless condition is not satisfied at
$\omega \sim 2 m_{\pi}$ at this higher incident energy.

So far, we have calculated  $N_{\rm eff}$ at a certain reaction $Q$ value
corresponding to each $\sigma$ eigenenergy, and treated it as a constant
in the energy spectrum.   This means that we have neglected
the finite size effects of the nucleus since the momentum transfer varies
according to the reaction $Q$ value and the nuclear form factor can
change the spectrum shape.
To see the effect
clearly, we show the contribution from the $[(5s)^{\sigma} \otimes
(2p_{1/2})^{neutron}]$
configuration $N_{\rm eff} \rho_{nl}(\omega)$ as a
function of
the kinetic energy of the emitted particle
T$_f$ in Fig.6 for the (d,t) reaction at T$_d$=4 GeV with
$C$=0.4.
The dotted line indicates the result without the nuclear size effects,
where the $N_{\rm eff}$ is calculated at the $\sigma$ eigenenergy and is
treated as a constant. The solid line is the result with the $N_{\rm eff}$
calculated with the appropriate momentum transfer for each T$_{f}$.  We
found that the nuclear form factor effects could have stronger energy
dependence than those of $\sigma$ meson spectral function in cases with
relatively larger eigenenergies $\omega_{nl}$.
We show in Fig.7 the total spectral functions $\rho^{(d,t)}_{\sigma -Pb}$
and $\rho^{(d,^{3}He)}_{\sigma -Pb}$ when we include the nuclear form factor
effects at T$_d$=1.5 GeV in the $C$=0.2 and 0.4 cases.  At this energy,
since the recoilless condition is satisfied at T$_f
\sim$T$_i$-2m$_{\pi}$, the spectra are enhanced due to the energy
dependence of the $N_{\rm eff}$.  Because of the enhancement of the
contributions from shallow $\sigma$ states around this region, the sharp
peak is relatively suppressed and we can only see the peak for the
(d,$^3$He) reactions.

To evaluate the magnitude of the
 cross section in the effective number approach, we need
the data of the cross section for the elementary process
d + n $\rightarrow$ t + $\sigma$ $\rightarrow$ t + $2 \pi$
and d + p $\rightarrow$ $^{3}$He + $\sigma$ $\rightarrow$ $^3$He + $2 \pi$.
   Since the data are not available at present, 
we estimate the elementary cross section from the 
$d+p\rightarrow$ $^{3}{\rm He}+X$ spectra in this energy region. We can find the $2\pi$
contributions fitted to the data in ref.\cite{ref:Bana}, which are
expected to include '$\sigma$' contributions. We roughly estimate the
elementary cross section to be $20 \mu b$/sr by assuming the $\sigma$
contribution is around $\frac{d^2\sigma}{dp_{\rm He}d\Omega_{\rm
He}}\sim 100\left(\frac{\mu{\rm b}}{{\rm sr\cdot GeV}/c}\right)$ in the figures in
ref.\cite{ref:Bana} and by multiplying the momentum interval $\Delta
p_{\rm He}=0.2 {\rm GeV}/c$ which corresponds to the missing mass of
$X$ $m_X=400\sim600{\rm MeV}$.
Using this value we can roughly estimate the order of the double
differential cross section of the emitted triton after the $\sigma$ bound
state formation
using the following expression;
\begin{equation}
\frac{d^2 \sigma}{d \Omega_{f} dE_{f}} = \left( \frac{d \sigma}{d \Omega}
\right) 2 m_{\sigma} \rho^{(d,{f})}_{\sigma-Pb} (\omega),
\label{eqn:dtcross}
\end{equation}
\noindent
where $f$ indicates the $t$ and/or $^3$He.  We remark that
 (\ref{eqn:dtcross})
 can be reduced to the same formula used in Ref. \cite{umemoto00}
with non-relativistic energies by identifying
$\Gamma= - {\rm Im}\Sigma /m_{\sigma}$,
$E_{\sigma}=(\omega^2 - m_{\sigma}^2)/2m_{\sigma}$
and
$E_{nl}=(\omega_{nl}^2 - m_{\sigma}^2)/2m_{\sigma}$, where $\Gamma$,
$E_{\sigma}$ and $E_{nl}$ are the $\sigma$ width, non-relativistic
$\sigma$ energy induced by the reaction and non-relativistic $\sigma$
eigenenergy of the ($nl$) state, respectively.  From Eq.
(\ref{eqn:dtcross}) we can estimate the cross section using the spectral
function as;
\begin{equation}
\frac{d^2 \sigma}{d \Omega_{f} dE_{f}} \left( \frac{\rm \mu b}{\rm sr \cdot GeV}
 \right)
\sim 20 \left( \frac{\rm \mu b \cdot GeV }{\rm sr} \right) \cdot \rho \left(
\frac{1}{\rm GeV^2} \right).
\end{equation}
\noindent
Using the results shown in Figs. 7, the cross sections are estimated to be
around 4 $\sim$ 10 [$\mu$b/sr/GeV] at the dominant
peak in the spectrum.
The background cross section is roughly estimated to be around $3
[{\mu{\rm b/sr/GeV}}]$ for light nuclear target cases in the first
reference of \cite{hayano99}.

Finally in order to discuss the sensitivities of the obtained results to
the theoretical model adopted, we compare our $\sigma$ mass and width to
the those obtained in ref. \cite{ref:0112048}. In Fig.\ref{fig:8}, we
show the $\sigma$ meson mass and width in nuclear matter as a function
of the nuclear density.
We can see the $\sigma$ mass is smaller for higher nuclear density for
all cases shown in Fig.\ref{fig:8}. In our model the $C$ parameter
determine the strength of the real part of the $\sigma$-Nucleus optical
potential, namely $\sigma$ mass in medium. We find that our results with
$C=0.2\sim0.3$ provide the similar strength of the real part of the
optical potential to that in ref.\cite{ref:0112048}.

As for imaginary part, the half widths of our model shown in
Fig. \ref{fig:8} have different $\rho$ dependence from
ref.\cite{ref:0112048}. In our model, the $\rho$ dependence of the
$\sigma$ half width is due to the $m_\sigma$ dependence of the
$\sigma\rightarrow \pi\pi$ decay width and, thus, the half width is $0$
for nuclear densities where $m_\sigma\le2m_\pi$.
In the model in ref.\cite{ref:0112048}, other channels, 
$\sigma\rightarrow\pi+ph$ and $\sigma\rightarrow2ph$, are implemented and
the half width shows the different $\rho$ dependence.
We can see from the figure that in the model of ref. \cite{ref:0112048}
the half-width is larger than around 100 MeV for $\rho=0\sim 1.4\rho_0$
and the narrow bound states will not exist.

We would like to mention that it is also instructive to compare our
spectral functions in Fig. 3(b) with those by Z. Aouissat {\it et al.} in
Fig. 2 in ref. \cite{shuck}, where the {\it ph} contribution is included.

\section{Summary}

We have studied the properties of the $\sigma$ meson
 at finite
density through the formation of the $\sigma$ mesic nuclei in the
  (d,t) and (d,$^3$He) reactions.
 This is related to the physics of
 partial restoration of the chiral symmetry at finite
 baryon density.

 The $\sigma$
meson embedded in nuclear matter is treated on the basis of
 the SU(2) linear $\sigma$-model
 in the one-loop approximation. By solving the Klein-Gordon equation for
 $\sigma$ in the local density approximation,
we  found that the deeply bound $\sigma$ states for heavy nuclei have
significantly smaller decay widths than those in vacuum due to
the $\pi\pi$ phase space suppression.
 By evaluating the spectral function of the $\sigma$ bound states
 in Pb,  a significant enhancement near the 2$\pi$ threshold
 is observed in spite of the large degeneracy factor for
 the shallower bound states.

 We have also evaluated the expected spectral function in the
(d,t) and (d,$^3$He) reactions using the effective number approach.
 The  ``shape'' of the spectral
 function becomes insensitive to the change of the
 chiral condensate in  nuclei because $N_{\rm eff}$ has
 large weight on the shallow states in these reactions.
 However, the absolute magnitude of the spectral function
 is sensitive to the change of the
 chiral condensate in nuclei.
 If we chose  T$_d$=1.5 GeV, the recoilless kinematics is met
 for $\omega \sim 2m_{\pi}$. In this case,
 the sharp peak structure near the 2$\pi$ threshold for $C=0.4$
 could be seen for the (d,$^3$He) reactions.
 By assuming the elementary cross sections to be 20 [$\mu$b/sr],
  the absolute value of the cross section to be 4 $\sim$ 10 [$\mu$b/sr/GeV].
 In order to observe the softening spectral function of
the $\sigma$-channel more clearly,
 other  reactions which can enhance the 
 contribution of  the deeply bound $\sigma$ states
 should be also considered.
 The evaluation of the background from e.g.
$\Delta$ processes, which will have a peak around 2$m_\pi$ excitation
energy region, is also necessary for quantitative study.

%
%
%
%

\section{Acknowledgment}

S. H. and H. N. acknowledge the hospitality of Department of
Physics of Tokyo Institute of Technology, where the part of this work
was carried out. S. H. acknowledges the stimulating discussions with Dr.
D. Jido.
S. H. was partially supported by the Grants-in-Aid of the Japanese
Ministry of Education, Culture, Sports, Science and Technology (No.
11694082 and No. 14540268).
 T. H. and T.K. were partially supported by the Grants-in-Aid of
the Japanese Ministry of Education, Science and Culture
(No. 12640263 and 12640296).










\pagebreak

\newpage

\begin{table}[h]
\caption{Parameters used in the present calculations and $\sigma$ meson
 mass in vacuum $m_\sigma^{\rm free}$ obtained in this model are shown \cite{CH98}.}
\vspace{3mm}
\centerline{
\begin{tabular}{|ccccc||c|}
\hline
$\lambda $                     &
$\kappa$ (MeV)                 &
$\sigma_0$ (MeV)               &
$m_{\sigma}$ (MeV)             &
 $m_{\pi}$ (MeV)               &
$m_{\sigma}^{\rm free}$ (MeV)  \\ \hline
 73.0  & 255  & 90.7 & 469 & 140 & 550     \\
\hline
\end{tabular}
}
\end{table}

\vspace{1.cm}
\begin{figure}
\begin{center}
\leavevmode
\epsfysize=7cm
\epsffile{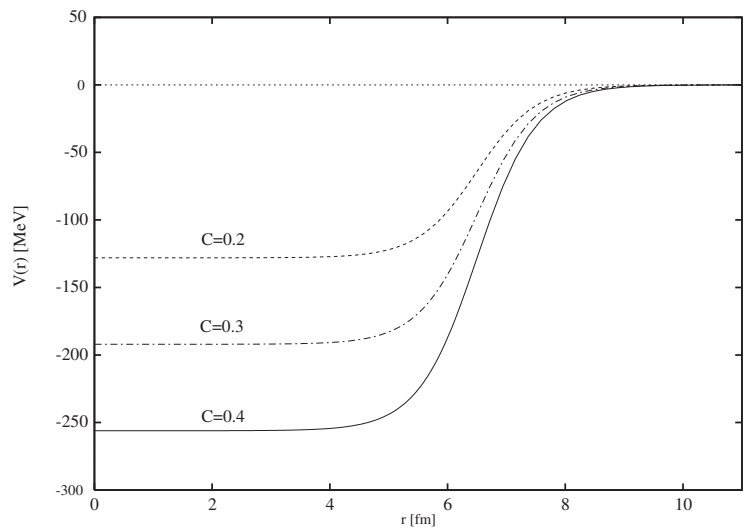}
\caption{The real potential for $\sigma$ meson inside the $^{208}$Pb nucleus
defined as;
$V(r) \equiv \Sigma_{MF}(\rho(r)) / 2m_{\sigma}$.
The density is assumed to be the Woods-Saxon form with 6.5 [fm] and 0.5
[fm]
radius and diffuseness parameters, respectively.  Each line indicates
C=0.2 (dotted line), 0.3 (dot-dashed line) and
0.4 (solid line) case, respectively.}
\label{fig:1}
\end{center}
\end{figure}
\begin{figure}
\begin{center}
\leavevmode
\makebox{
\epsfysize=5cm
\epsffile{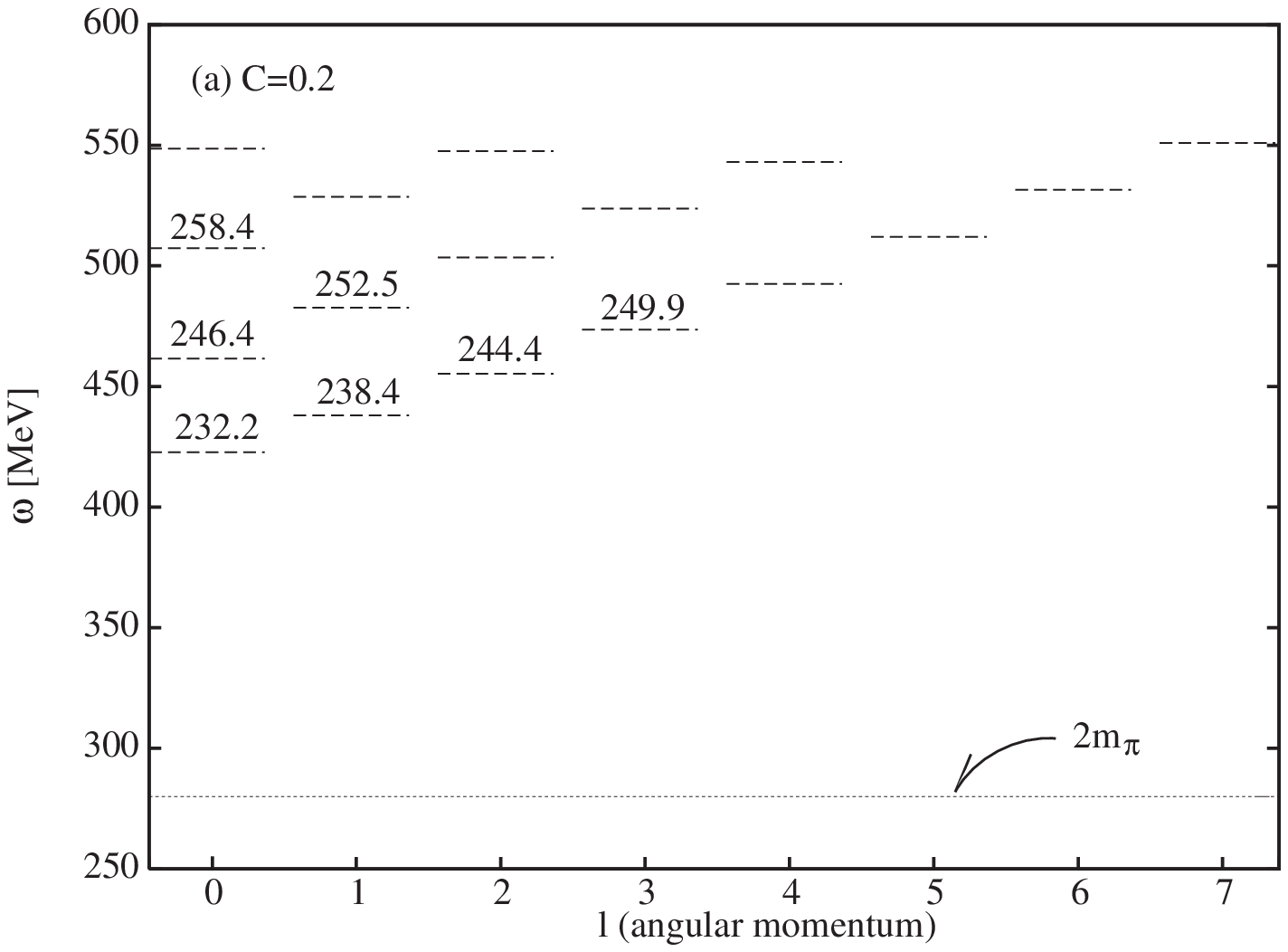}
\epsfysize=5cm
\epsffile{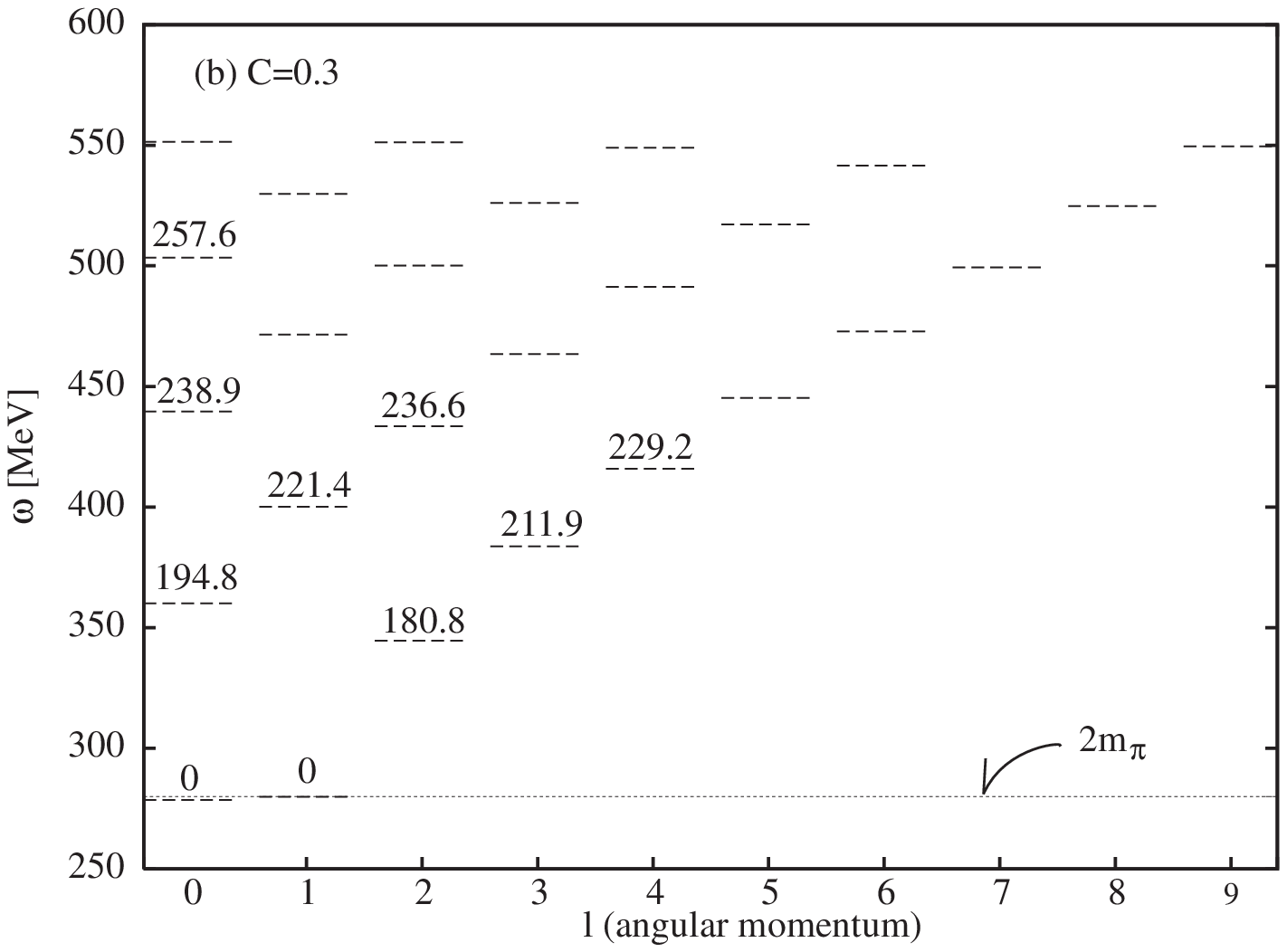}}\\
\vspace{5mm}
\epsfysize=5cm
\epsffile{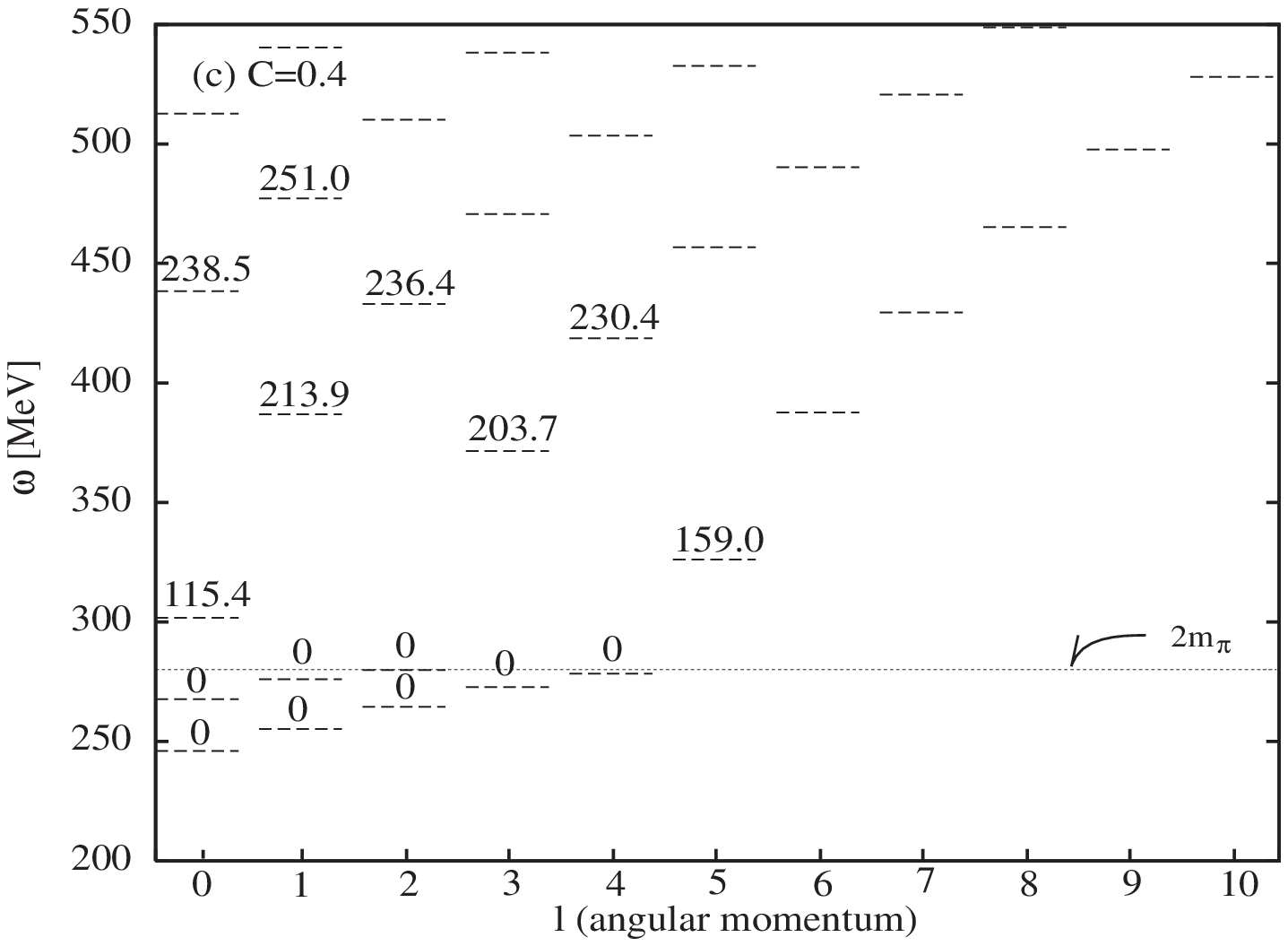}

\caption{
Calculated eigenenergies of the bound $\sigma$ in $^{208}$Pb with (a) C=0.2,
(b)
C=0.3 and
(c) C=0.4. The $l$ is the orbital angular
momentum of the $\sigma$.  Widths are shown by numbers
for the low lying states in unit of MeV.  The $ 2 m_{\pi} $ threshold is
shown by the dotted line. The states below the threshold do not have
widths due to the $\pi \pi$ decay.}
\label{fig:2}
\end{center}
\end{figure}
\begin{figure}
\begin{center}
\leavevmode
\epsfysize=13cm
\epsffile{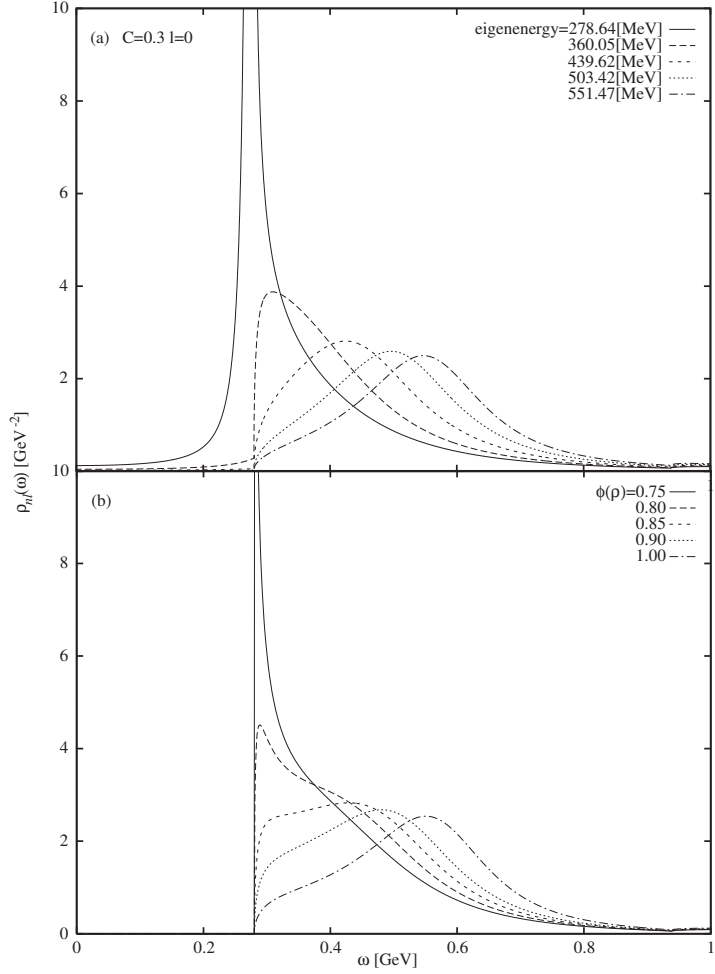}
\caption{(a) The spectral function $\rho_{nl} (\omega)$ of the $\sigma$ bound
states with $l$=0 for C=0.3 case. 
We added 5MeV
to
the imaginary part of the self-energy.
(b) Spectral function of the $\sigma$ in nuclear medium for several values
of $\Phi =
\langle\sigma\rangle_{\rho} / \sigma_0$ \cite{hatsuda99}.
}
\label{fig:3}
\end{center}
\end{figure}
\begin{figure}
\begin{center}
\leavevmode
\epsfysize=7cm
\epsffile{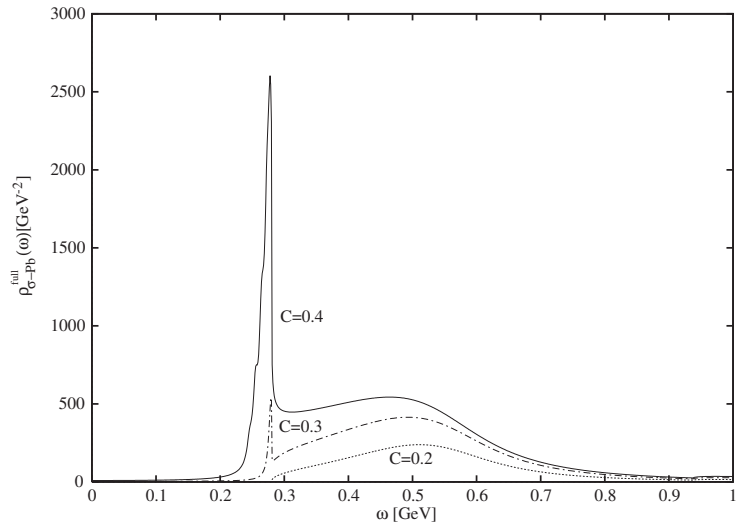}
\caption{
Full spectral functions of the $\sigma$ in Pb, which include
all bound states contributions, are
shown for C=0.2 (dotted line), 0.3 (dot-dashed line) and 0.4 (solid
line) cases.}
\label{fig:4}
\end{center}
\end{figure}
\begin{figure}
\begin{center}
\leavevmode
\epsfysize=10cm
\epsffile{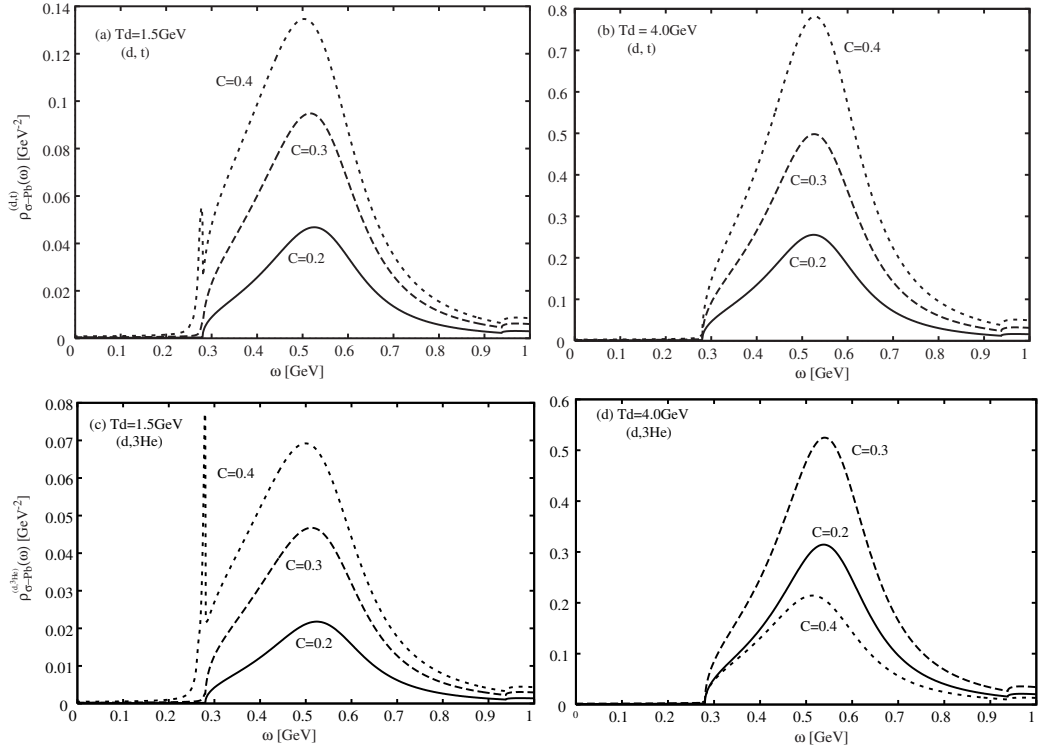}
\caption{
Total spectral function obtained with the effective numbers for (d,t)
reactions as the
relative strength of the each $\sigma$ eigenstate at (a) T$_d$=1.5 GeV
and (b) 4GeV, and for the (d,$^3$He) reaction (c) T$_d$=1.5 GeV
and (d) 4GeV.  Each line indicates
C=0.2 (solid line), 0.3 (dashed line) and
0.4 (dotted line) case, respectively.}
\label{fig:5}
\end{center}
\end{figure}

\begin{figure}
\begin{center}
\leavevmode
\epsfysize=6cm
\epsffile{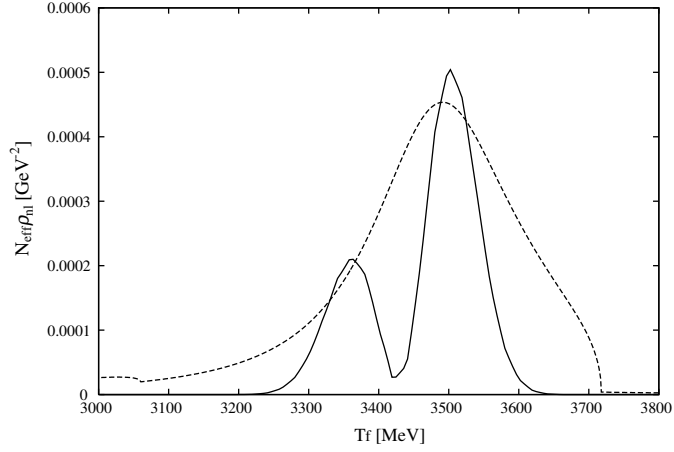}
\caption{The N$_{eff} \rho_{nl}(\omega)$ for the $[(5s)^{\sigma} \otimes
(2p_{1/2})^{neutron}]$ configuration is plotted as a function of
the kinetic energy of the emitted triton
T$_f$
for the (d,t) reaction at T$_d$=4GeV with $C$=0.4.  The solid line
indicates the spectrum with the energy dependent N$_{eff}$ and the dashed
line with the constant N$_{eff}$ calculated at the $\sigma$ eigenenergy,
$\omega$=512.64 MeV. }
\label{fig:6}
\end{center}
\end{figure}

\begin{figure}
\begin{center}
\leavevmode
\epsfysize=6cm
\epsffile{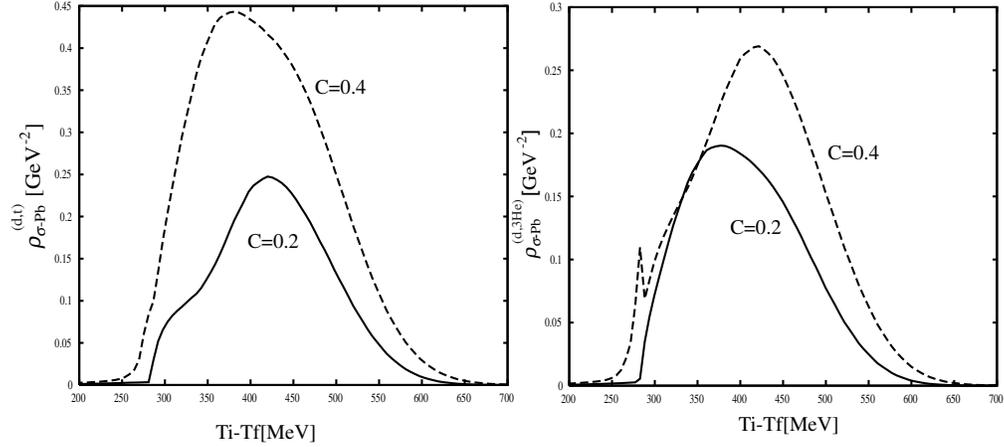}
\caption{Total spectral function obtained with the energy dependent
effective numbers for the (d,t) and (d,$^3$He) reactions at T$_d$=1.5 GeV.}
\label{fig:7}
\end{center}
\end{figure}

\begin{figure}
\begin{center}
\leavevmode
\epsfysize=10cm
\epsffile{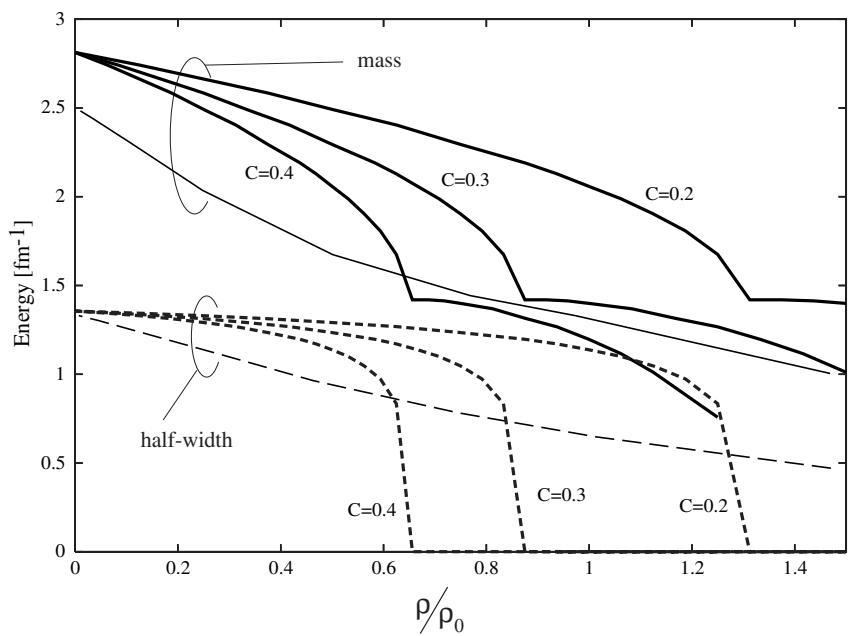}
\caption{Calculated $\sigma$ meson mass (solid lines) and half-width
 (dashed lines) in nuclear matter are shown as a
 function of the nuclear density.
Present results are shown as thick lines for $C=0.2, 0.3$ and $0.4$
 cases.
The results obtained in ref. \cite{ref:0112048} are shown as thin lines
for comparison.
}

\label{fig:8}
\end{center}
\end{figure}

\end{document}